\documentclass[prl,twocolumn,showpacs,preprintnumbers,amsmath,amssymb,floatfix]{revtex4}

\usepackage{subfigure,graphicx,epsfig,amsmath,amsfonts,amssymb,xcolor,slashed,ulem,multirow}

\newcommand{\be}{\begin{equation}}
\newcommand{\ee}{\end{equation}}
\newcommand{\ba}{\begin{eqnarray}}
\newcommand{\ea}{\end{eqnarray}}
\newcommand{\nn}{\nonumber}
\newcommand{\mev}{\textrm{ MeV}}

\begin{document}

\title{Charm-beauty meson bound states from $B(B^*)D(D^*)$ and
$B(B^*)\bar D(\bar D^*)$  interaction}

\author{ S.~Sakai$^1$, L.~Roca$^2$, and E.~Oset$^1$}

\affiliation{$1)$ Departamento de
F\'{\i}sica Te\'orica and IFIC, Centro Mixto Universidad de
Valencia-CSIC Institutos de Investigaci\'on de Paterna, Aptdo.
22085, 46071 Valencia, Spain\\
$2)$ Departamento de F\'isica, Universidad de Murcia, E-30100 Murcia, Spain}

\date{\today}

\begin{abstract} 
 We evaluate the s-wave interaction of pseudoscalar and vector mesons
 with both charm and beauty to investigate the possible existence of
 molecular  $BD$, $B^*D$, $BD^*$, $B^*D^*$, $B\bar D$, $B^*\bar D$,
 $B\bar D^*$ or $B^* \bar D^*$ meson states. The scattering amplitude is
 obtained implementing unitarity starting from a tree level potential
 accounting for the dominant vector meson exchange. The diagrams are
 evaluated using suitable extensions to the heavy flavor sector of the
 hidden gauge symmetry Lagrangians involving vector and pseudoscalar
 mesons{, respecting heavy quark spin symmetry}. 
 We obtain bound states
 at energies
 above 7~GeV for 
 $BD$  ($J^P=0^+$), $B^*D$ ($1^+$), $BD^*$ ($1^+$) and $B^*D^*$
 ($0^+$, $1^+$, $2^+$), all in isospin 0.
 For $B\bar D$  ($0^+$), $B^*\bar D$ ($1^+$), $B\bar D^*$ ($1^+$) and $B^*\bar D^*$
 ($0^+$, $1^+$, $2^+$) we also find similar bound states in $I=0$,
 but much less bound, which
 would correspond to exotic meson states with $\bar b$ and $\bar c$
 quarks, and for the $I=1$ we find a repulsive interaction. We also
 evaluate the scattering lengths in all cases, which can be tested
 in current investigations of lattice QCD.
\end{abstract}

\maketitle

\section{Introduction}
The present situation of the mesons with one quark of type $b$ and an
antiquark of type $c$,  $B_c$ mesons, is still at an early
beginning. There are just two states reported in the PDG (Particle data
Book) \cite{pdg}, the $B_c(6275)$ and the $B_c(2S)(6842)$. This
contrasts with the situation in the bottom
strange sector, where we have the states $B_s(5367)$, $B_s^*(5415)$,
$B_{s1}(5830)$, $B_{s2}(5840)$, $B_{s2}(5850)$ and in the charm strange
sector where there are already ten $D_s$ states reported, with an
average separation between the masses of about 100 MeV. By contrast, the
only two $B_c$ states reported are separated by nearly 600~MeV.
Lattice QCD has also made a contribution to the heavy meson sector, investigating possible tetraquarks or molecular states 
\cite{Ikeda:2013vwa,Guerrieri:2014nxa,Bicudo:2015vta,Bicudo:2015kna,Francis:2016hui}, however, none of them deals with the $BD$ quantum numbers.

 It is
clear that many states are missing which most hopefully will be
discovered in coming years. An idea of the advance made in time is the
addition of three new $D_s$ states since the 2008 edition of the PDG
\cite{pdg08} and one $B_c$ state. Yet, the advent of LHCb has made the
prognosis brighter, one recent example being the determination of five
new $\Omega_c$ states \cite{lhcb}.

  Although some of the states expected should correspond approximately to the $q \bar q$ standard structure of the mesons, the irruption of so many XYZ states \cite{olsenxyz}, which do not fit into the traditional $q \bar q$ picture, motivated a large number of theoretical studies that go beyond this picture, invoking especial quark configurations 
\cite{Segovia:2013wma,Vijande:2014cfa}, tetraquarks  
\cite{Wu:2017weo,Chen:2016ont,Wu:2016gas,polotetra} or meson meson molecules 
\cite{Ortega:2012rs,Yang:2017prf,Kolomeitsev:2003ac,Hofmann:2003je,Gamermann:2006nm,
Molina:2009ct,Ozpineci:2013qza,Dias:2014pva,Xiao:2013yca,Albaladejo:2016ztm,Hidalgo-Duque:2013pva,Guo:2013sya,Sun:2012sy,Sun:2011uh,Ohkoda:2012hv,Lee:2009hy,Dong:2017gaw,Liu:2017mrh,He:2014nya}. Mixtures
of charmonium states and molecules have also been investigated
\cite{Cincioglu:2016fkm} and methods to disentangle the nature of the
states have been suggested
\cite{Zhao:2014gqa,Cleven:2015era,Ma:2014ofa,liangmolina,branz}. Reviews on
these issues are available in Refs.~\cite{xliuxyz,Chen:2016qju,Swanson:2015wgq}.

In the present work we take the case of the interaction of $B (B^*)$
and $D (D^*)$ mesons, an also the corresponding cases with $\bar D (
\bar D^*)$. Given the analogy of the $B $ meson with a $K$ meson, the
states we study have an analogy with the $DK$, $D K^*$, $D^* K$
interactions.
According to {Ref}.~\cite{Gamermann:2006nm} the $DK$ channel is
the main building block of the $D_{s0}^*(2317)$, something that is
corroborated by the analysis of lattice QCD results in the light-heavy
sector \cite{sasa}. Similarly, the $D^* K$ component appears as the
main building block of the $D_{s1}(2460)$ in Ref.~\cite{daniaxial}, which
is again corroborated by the lattice QCD study of Ref.~\cite{sasa}. And in
Ref.~\cite{daniaxial} one also finds that the $D_{s1}(2536)$ resonance is
mostly formed from the $D K^*$ component.
{Similarly the $D^*K^*$ interaction appears as the main building block
of the $D_{s2}^*(2573)$ in Ref.~\cite{branzbound}.}
The $D^* \bar D^*$ interaction
is also studied in Ref.~\cite{Molina:2009ct} and bound states are reported
there. In view of that,
it is reasonable to expect bound states of the $B (B^*) D (D^*)$
systems,
which we study in the present work.
The formalism that we use is the
local hidden gauge approach
\cite{Bando:1984ej,Bando:1987br,hidden4,Nagahiro:2008cv}, which combines
pseudoscalar and vector mesons, properly extended to the heavy quark sector \cite{Molina:2009ct}.
The interaction stems from the exchange
of vector mesons between the interacting mesons, and in the limit of
small momentum transfers this gives rise to the chiral Lagrangians
in the light quark sector. An
example for the interaction of vector-pseudoscalar is given in
Ref.~\cite{Nagahiro:2008cv} where it is shown that it gives rise to the
chiral Lagrangian of Ref.~\cite{Birse:1996hd}.
{It is also interesting to mention that the exchange of light vector
mesons between hadrons involving heavy quarks respects heavy quark
symmetry \cite{hqss} as shown in Refs.~\cite{Xiao:2013yca,lianghqss}.}

  We find that all the four systems lead to bound states in $I=0$, and
  in the case of $B^* D^*$ there are three spin states, degenerate in
  energy within the model. We also study the $B (B^*) \bar D (\bar D^*)$
  systems and here we find that there is attractive interaction in $I=0$
  and repulsive interaction in $I=1$. In the case of $I=0$ we also find
  bound states which would be neatly exotic since they contain a $\bar
  b$ and a $\bar c$ quark. 

  We also evaluate the scattering lengths in all cases, since lattice QCD calculations start providing such observables in the heavy quark sector 
\cite{Liu:2008rza,Mohler:2012na,Lang:2014yfa,Mohler:2013rwa} and they have proved useful to constraint parameters in effective theories \cite{Yao:2015qia}.

\section{Formalism}

\subsection{Elementary interaction via vector-meson exchange}

One of the most successful realization{s} of chiral symmetry  when vector mesons are involved is the hidden gauge symmetry (HGS) formalism \cite{Bando:1984ej,Bando:1987br,hidden4,Nagahiro:2008cv}, where the vector meson fields are
gauge bosons of a hidden local symmetry transforming inhomogeneously, and is the most natural way to account for vector meson dominance.
The extension of the HGS approach  to the charm \cite{Molina:2009ct,branzbound} and beauty 
quark sector \cite{Ozpineci:2013qza,Dias:2014pva,branzbound} 
  has turned out to be very useful to deal with meson-meson and
  meson-baryon interactions involving hidden and open charm and beauty
  mesons and baryons. Furthermore it has been also shown in
  Refs.~\cite{Xiao:2013yca,lianghqss} that HGS respects the heavy quark
  spin symmetry (HQSS), which is the symmetry of QCD by means of which
  for heavy quarks their interaction is independent of the spin.

\begin{figure}[tbp]
     \centering
     \includegraphics[width=.8\linewidth]{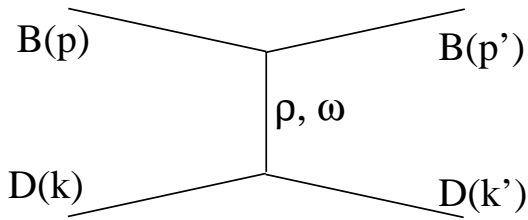}
    \caption{$BD$ interaction via vector-meson exchange}
 \label{fig:diagBD}
\end{figure}

Let us illustrate the formalism with the $BD$ channel, since the other ones are analogous and the peculiarities of the different channels will be pointed out when necessary. 
In the HGS approach, the  $BD$ interaction would proceed through the
exchange of a vector meson, as depicted in Fig.~\ref{fig:diagBD}.
 The exchange of light vector mesons, $\rho$ and $\omega$ are by far the dominant ones since the vector propagator contributes as $1/m_V^2$ and thus possible exchange of vector mesons containing heavy flavors are very suppressed.
We thus need the vector-pseudoscalar-pseudoscalar ($VPP$)  Lagrangian
\begin{equation}
\mathcal{L}_{VPP}=-ig\langle V^{\mu}[P,\partial_{\mu}P]\rangle\ ,
\label{eq:VPPlag}
\end{equation}
where $g=M_{V}/2f$, $M_V$ is the vector meson mass, with $f=93$~MeV the pion decay constant, and $\langle\cdots\rangle$ stands for $SU(4)$ trace.
Since the strange quark is not needed in the present work, it is sufficient to write the $P$ matrix in Eq.~(\ref{eq:VPPlag}) in
 SU(4) ($u$, $d$, $c$ and $b$ flavors) and is given by
\begin{equation}
P=\left(
\begin{array}{cccc}
0 & 0 & B^+&\bar{D}^0\\
0 & 0 & B^{0}&D^-\\
B^{-} & \bar{B}^{0}&0&B^-_c\\
D^0&D^+&B^+_c&0
\end{array}
\right)\ ,
\label{eq:pfields}
\end{equation}
where we do not show the light pseudoscalars which are not relevant for the present work.
Analogously, for the vector mesons we have
\begin{equation}
V=\left(
\begin{array}{cccc}
\frac{\omega}{\sqrt{2}}+\frac{\rho^0}{\sqrt{2}} & \rho^+ & B^{*+}&\bar{D}^{*0}\\
\rho^- &\frac{\omega}{\sqrt{2}}-\frac{\rho^0}{\sqrt{2}} & B^{*0}&D^{*-}\\
B^{*-} & \bar{B}^{*0} &0&B^{*-}_c\\
D^{*0}&D^{*+}&B^{*+}_c&0
\end{array}
\right) .
\label{eq:vfields}
\end{equation}

\begin{figure}[tbp]
     \centering
     \includegraphics[width=.98\linewidth]{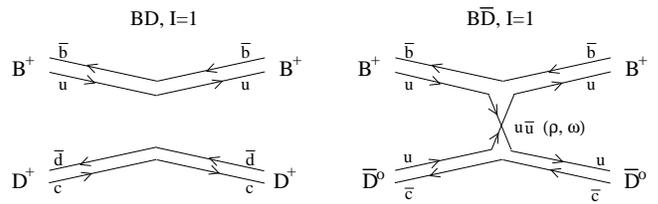}
    \caption{Elementary isospin  $I=1$ $BD$ and $B\bar D$  diagrams at
 quark level which show why
 the interaction is
 zero for $BD$  and not for $B\bar D$.}
 \label{fig:diagBDI1}
\end{figure}

Since  both $D$ and $B$ are isospin $I=1/2$ states, the total $BD$
isospin can be 0 and 1. However, the $I=1$ interaction is very small
since, from the above Lagrangians, it can be obtained that the $\rho$
and $\omega$ exchange contributions for this isospin channel have
different   sign and they cancel among themselves, up to the small
difference between the masses squared of the $\rho$ and $\omega$. This
is not the case for the $B\bar D$ interaction in $I=1$, where $\rho$ and
$\omega$ contributions have the same sign. This can also be understood
at the quark level by looking at the diagrams of
Fig.~\ref{fig:diagBDI1}, where one can see that for the $BD$ case in
$I=1$ it is not possible to exchange a vector meson at first order,
while for $B\bar D$ it is allowed via $u\bar u$ exchange.

It might look that we are making use of $SU(4)$ symmetry by using
Eqs.~(\ref{eq:pfields}) and (\ref{eq:vfields}), but actually writing the Lagrangian in this form is only a practical way to obtain the couplings of the heavy mesons to the light vectors, that we can also obtain in a very simple picture where the heavy quarks are spectators, in the spirit of the heavy quark formalism, and we are only making use of $SU(2)$ symmetry. Indeed, we can write the $\rho^0$, $\omega$, sources ($\phi$ is not present in our case) as 
\ba
&&g\frac{1}{\sqrt{2}}(u\bar u-d \bar d)\ , \qquad \textrm{for $\rho^0$ exchange},\nn\\
&&g\frac{1}{\sqrt{2}}(u\bar u+d \bar d)\ , \qquad \textrm{for $\omega$ exchange},
\label{eq:rhowquarks}
\ea
and taking into account the vector type coupling we would have the operator 
\be
-g\frac{1}{\sqrt{2}}\left((u\partial_\mu \bar u-\partial_\mu u \,\bar u)
-(d\partial_\mu \bar d-\partial_\mu d \, \bar d)\right).
\ee
Let us study, as an example, the cases of $B^0 B^0 \rho^0$, $D^+ D^+ \rho^0$ and the other cases follow directly from them. The heavy mesons are $B^0=\bar b d$, $D^+=c\bar d$ and, since the heavy quarks are spectators, we have the matrix elements
\ba
&&-\langle \bar b d|g\frac{1}{\sqrt{2}}\left(
 (u\partial_\mu \bar u-\partial_\mu u \,\bar u )
-(d\partial_\mu \bar d-\partial_\mu d \, \bar d)\right)
|\bar bd\rangle \nn\\
&&=-g\frac{1}{\sqrt{2}}(-ip_\mu-ip'_\mu)
\ea
for $B^0 B^0 \rho^0$ and 
\ba
&&-\langle  c \bar d|g\frac{1}{\sqrt{2}}\left(
 (u\partial_\mu \bar u-\partial_\mu u \,\bar u )
-(d\partial_\mu \bar d-\partial_\mu d \, \bar d)\right)
|c \bar d\rangle \nn\\
&&=-g\frac{1}{\sqrt{2}}(ip_\mu+ip'_\mu)
\ea
for $D^+ D^+ \rho^0$,
where $p$ ($p'$) is the light quark initial (final) momentum.
In the limit of $B$  at rest, $p_\mu+p'_\mu$ will become $2 m_q \delta_{\mu 0}$, with $m_q$ the mass of the light quark.  Let us compare this with the coupling of $K^0 K^0 \rho^0$ ($K^0= \bar s d$), which in the limit of the $K^0$ at rest gives us the same contribution $2 m_q \delta_{\mu 0}$. This means that in the spectator picture for the $b$ or $s$ quarks the matrix element for $B^0 B^0 \rho^0$, $K^0 K^0 \rho^0$, are the same at the microscopic quark level. However, when we write the amplitudes at macroscopic hadron level, we must take into account that the S-matrix has the field normalization factors $\frac{1}{\sqrt{2 M_H}}$ 
\cite{mandl}
for each external hadron ($H$) (see Eqs.~(14)-(16) of ref.~\cite{lianghqss}).
Hence, at the macroscopic level we would have at threshold
\be
\frac{t_{B^0 B^0 \rho^0}}{t_{K^0 K^0 \rho^0}}=\frac{M_B}{M_K}.
\ee
Since 
\be
-i t_{{K^0 K^0 \rho^0},\mu}=-g\frac{1}{\sqrt{2}}(-iM_K-iM_K)\delta_{\mu 0},
\ee
then
\be
-i t_{{B^0 B^0 \rho^0},\mu}=-g\frac{1}{\sqrt{2}}(-iM_B-iM_B)\delta_{\mu 0},
\ee
and in covariant form
\be
-i t_{{B^0 B^0 \rho^0},\mu}=-g\frac{1}{\sqrt{2}}(-ip_\mu-i p'_\mu),
\ee
and this is what we get straightforwardly from the use of the Lagrangian of Eq.~(\ref{eq:VPPlag}).
It is also interesting to see that the relative sign between 
$B^0 B^0 \rho^0$ and $D^+ D^+ \rho^0$ comes because in $B^0$ we have a $d$ quark and in  $D^+$ we have a $\bar d$ quark and we have the operator $q\partial_\mu \bar q-\partial_\mu q\, \bar q$. One can immediatly see that if we consider $B^0 B^0 \omega$ and $D^+ D^+ \omega$ couplings, using Eq.~(\ref{eq:rhowquarks}), we would get opposite sign to the cases 
$B^0 B^0 \rho^0$ and $D^+ D^+ \rho^0$.
One can see that all other cases follow automatically and one obtains exactly the same results as with the Lagrangian of Eq.~(\ref{eq:VPPlag}) with $SU(4)$ matrices of Eqs.~(\ref{eq:pfields}) and (\ref{eq:vfields}).
It is interesting to see that the same arguments used for the 
$D^* \to D\pi$ and $B^* \to B\pi$ coupling  \cite{lianghqss} lead to results in agreement with experiment and lattice QCD results respectively \cite{pdg,hqss,Flynn:2013kwa}.
\\

\begin{figure}[tbp]
     \centering
     \includegraphics[width=.98\linewidth]{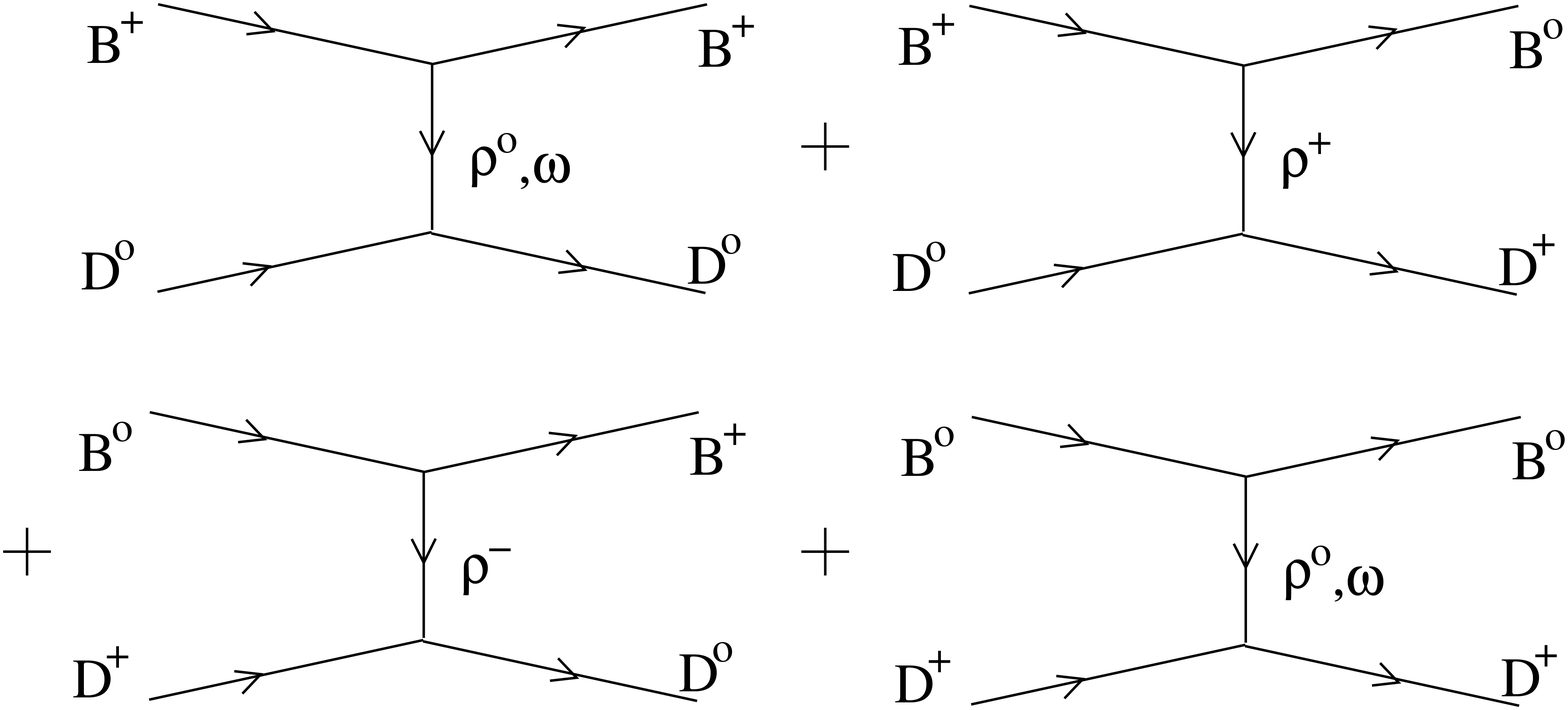}
    \caption{Vector meson exchange contribution for $BD$ interaction in isospin $I=0$.}
 \label{fig:BDcharge}
\end{figure}

In order to evaluate the $I=0$ $BD$ interaction we need the $I=0$
combination{, with the doublets $(B^+,B^0)$, $(D^+,-D^0)$},

\be
|BD\rangle^{(I=0)}=-\frac{1}{\sqrt{2}}\left(|B^+ D^0\rangle+
|B^0 D^+\rangle\right)
\ee
and therefore the $I=0$ amplitude can be written as

\begin{align}
t^{(I=0)}_{BD\to BD}=\frac{1}{2}\left(t_{B^+ D^0\to B^+ D^0}
+t_{B^+ D^0\to B^0 D^+} \right.\nn\\
\left. + t_{B^0 D^+\to B^+ D^0}
+t_{B^0 D^+\to B^0 D^+}\right)
\label{eq:tI0}
\end{align}
The amplitudes in the bracket in the previous equation account for  the diagrams in Fig.~\ref{fig:BDcharge}.

From the Lagrangian of Eq.~(\ref{eq:VPPlag}), the amplitude of
Eq.~(\ref{eq:tI0}) can be readily calculated, leading to
\be
 t^{(I=0)}_{BD\to BD}=-\frac{1}{2 f^2} (p+p')_{\mu}(k+k')^{\mu}
 \label{eq:tI0noproj}
\ee
where we have approximated $m_V^2=m_\rho^2\simeq
m_\omega^2$. In addition,  in the derivation of
Eq.~(\ref{eq:tI0noproj}), we have neglected the momentum transferred in
the vector meson propagator. (The correction due to this effect will be
taken into account below including form factors in the loop functions
that will appear in the unitarization procedure.)

There is still one more issue that should be discussed since the heavy
quark spectator hypothesis in the $BB\rho$ coupling has been challenged
by some other approaches.
Indeed evaluations using the Dyson-Schwinger equation \cite{bruno} or
QCD lattice simulations \cite{oka} lead to a value of
$g_{DD\rho}$ about twice as big as $g_{KK\rho}=g_{DD\rho}$ of
Ref.~\cite{haiden}, which we are using here in Eq.~(\ref{eq:VPPlag}).
However, the couplings go together with a form factor leading to a
stronger off-shell reduction than the one used here.
A detailed discussion of this issue and the added uncertainties to our
approach, extending the discussion to the $B$ sector, was done in
section VII of Ref.~\cite{Xiao:2013yca}, concluding that it added extra
uncertainties in the bindings, increasing the binding by an amount which
could be as large as 40\%.
Although our system here is different than those studied in \cite{Xiao:2013yca}, it still deals with heavy mesons, and the exercise done here serves to give an idea of possible uncertainties and a hint that those considerations might increase the binding that we get.

After projecting over s-wave, Eq.~(\ref{eq:tI0noproj}) reads
\be
 t^{(I=0,\textrm{ s-wave})}_{BD\to BD}=-\frac{1}{4 f^2}\left[ 
 3s-2(m_B^2+m_D^2)-\frac{(m_B^2-m_D^2)^2}{s}\right]
 \label{eq:tI0swave}
\ee

\begin{figure}[tbp]
 \centering
 \includegraphics[width=7.5cm]{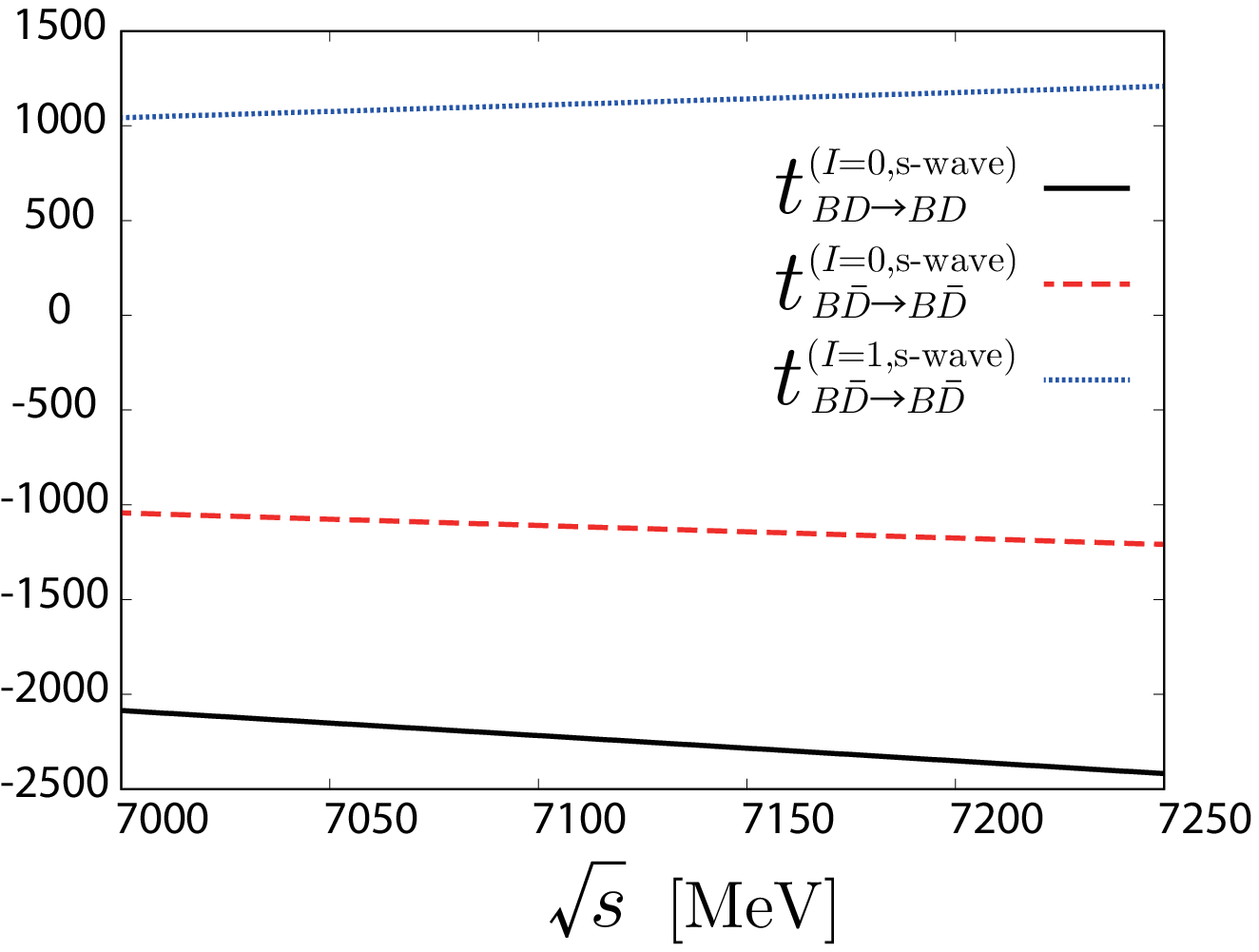}
 \caption{$t^{(I=0,\textrm{ s-wave})}_{BD\to BD}$, $t^{(I=0,\textrm{
 s-wave})}_{B\bar D\to B\bar D}$, and $t^{(I=1,\textrm{ s-wave})}_{B\bar
 D\to
 B\bar D}$ as functions of $\sqrt{s}$.
 These
 kernels
 are dimensionless.}
 \label{fig_vint}
\end{figure}

For the other channels we are considering in the present work, $B^*D$, $BD^*$, $B^*D^*$,
$B\bar D$, $B^*\bar D$, $B\bar D^*$ or $B^* \bar D^*$ the formalism is analogous to the $BD$ case with the following particular features and considerations:
\begin{itemize}
\item{$B^*D$ and $BD^*$:}
The $SU(4)$ matrices have the same  structure as in the previous case
     since the quark content of $B^*$ is the same as $B$, and $D^*$ the
     same as $D$. Furthermore in Ref.~\cite{Ozpineci:2013qza} it was
     justified by using HQSS that in the heavy sector the
     vector-pseudoscalar interaction is the same than
     pseudoscalar-pseudoscalar
     at leading order in the inverse of the heavy quark mass.
Therefore, the only difference with the $BD$ case is the vector character of the $B^*$ and $D^*$ which implies that, 
neglecting terms of order $q^2/m_V^2$ \cite{Oset:2009vf},
 one has to  add an $\vec \epsilon\cdot\vec \epsilon\,'$ factor in Eq.~(\ref{eq:tI0swave}) (where 
$\vec \epsilon(\vec \epsilon\,')$ is the initial(final) vector polarization vector) and the masses must be replaced by $m_{B^*}$ or $
m_{D^*}$ accordingly.

 \item{$B^*D^*$:}
      Again the flavor structure is the same and analogous arguments than before apply. In addition, a contact $VVVV$ term from the HGS Lagrangian
      $\mathcal{L}_{VVVV}=\frac{g^2}{2}\langle V^{\mu}V^{\nu}V^{\mu}V^{\nu}-V^{\nu}V^{\mu}V^{\mu}V^{\nu}\rangle$
      would be present but is subdominant \cite{pedro}
      and can thus be neglected. Furthermore, all four external
      particles are now vector mesons   and thus it turns out that we
      can use the same expression as  Eq.~(\ref{eq:tI0swave}) but adding
      a factor {$\vec \epsilon_{B^*}\cdot\vec \epsilon\,'_{B^*}\,
      \vec \epsilon_{D^*}\cdot\vec \epsilon\,'_{D^*}$} \cite{Molina:2008jw}
      and replacing $m_{B}$ and $m_{D}$ by $m_{B^*}$ and $m_{D^*}$.
 
 \item{$B\bar D$, $B^*\bar D$, $B\bar D^*$ or $B^* \bar D^*$:}
      We can analogously calculate the same interactions as before but substituting 
      $D$ and  $D^*$ by
      $\bar D$ and  $\bar D^*$. In this case we find attractive potential for $I=0$
      \be
      t^{(I=0,\textrm{ s-wave})}_{B\bar D\to B \bar D}=-\frac{1}{8 f^2}\left[ 
      3s-2(m_B^2+m_{ D}^2)-\frac{(m_B^2-m_{ D}^2)^2}{s}\right]
      \label{eq:tI0swavebar}
      \ee
      \noindent and repulsive for $I=1$:
      \be
      t^{(I=1,\textrm{ s-wave})}_{B\bar D\to B \bar D}= \frac{1}{8 f^2}\left[ 
      3s-2(m_B^2+m_{D}^2)-\frac{(m_B^2-m_{ D}^2)^2}{s}\right]
      \label{eq:tI1swavebar}
      \ee
      \noindent  and similarly for the vector meson cases substituting the corresponding masses.
\end{itemize}
{The kernels $t^{(I=0,\textrm{ s-wave})}_{BD\to BD}$, $t^{(I=0,\textrm{
s-wave})}_{B\bar D\to B \bar D}$, and $t^{(I=1,\textrm{ s-wave})}_{B\bar
D\to B \bar D}$ as functions of $\sqrt{s}$ are given in
Fig.~\ref{fig_vint}, which shows the attractive nature of the $BD$ and
$B\bar D$ $(I=0)$ interactions and the repulsive one of $B\bar D$
$(I=1)$.}

We can think of other elements that could be exchanged, apart from
the vector mesons considered.
A thorough investigation of other possible mechanisms was made in
Refs.~\cite{dias1,dias2} in the study of the $D\bar{D}^*$ interaction
with $I=1$ and its relationship to the $Z_c(3900)$ (or $Z_c(3885)$)
state in one case and in the study of $B\bar{B}^*$, $B^*\bar{B}^*$
interaction with $I=1$ and its relationship to the $Z_b(10610)$,
$Z_b(10650)$ states in the other case.
This was done because in these cases there is no light vector exchange
and then one could only exchange $J/\psi$ in one case and $\Upsilon$ in
the other, which made the vector exchange very small and gave chances to
other mechanisms to contribute.
One of the possible mechanisms was the exchange of two pions, uncorrelated
(non interacting) or correlated (interacting).
The case of two pion exchange with interacting pions gives rise to
``$\sigma$'' exchange in this picture, as was shown in Ref.~\cite{mizobe}.
The conclusion of Ref.~\cite{dias1} was that the two pion exchange still gave
a factor of four smaller contribution than the $D\bar{D}^*\rightarrow
\eta_c\rho$ or $D\bar{D}^*\rightarrow\pi J/\psi$ transitions that involve
a $D^*$ exchange.
Considering that the light vector exchange  potential gives an
$m_\rho^{-2}$ dependence rather than $m_{B^*}^{-2}$ in the
$D\bar{D}^*\rightarrow \eta_c\rho$ transitions, this gives a suppression
of a factor about 30
of the two pion exchange with respect to  the light vector exchange when
it is allowed, as in the present case.
Similar conclusions can be reached from the results in the $B$ sector
when one increases the $B^*$ exchange potential in terms like
$B\bar{B}^*\rightarrow \rho\Upsilon$ by the ratio $m_{B^*}^2/m_\rho^2$
to compare the two pion exchange with an allowed light vector
transition.

\subsection{Implementation of unitarity}

Using the techniques of the coupled channels unitary approach, exact unitarity can
be implemented into the $BD$ interaction, which can be carried out by
means of the Bethe-Salpeter equation: (equivalent to the $N/D$
\cite{Oller:1998zr,Oller:2000fj} or IAM
\cite{Dobado:1996ps,Oller:1998hw} methods)
\be
T=[1-VG]^{-1}V\, ,
\label{eq:T}
\ee
where $V$ is the potential, or kernel of the unitarization procedure, provided by Eq.~(\ref{eq:tI0swave}) and $G$ is the $BD$ loop function:
\begin{equation}
G=i\int\frac{d^4q}{(2\pi)^4}\frac{1}{q^2-m_B^2+i\epsilon}\frac{1}{(q-P)^2-m_D^2+i\epsilon}\ ,
\label{eq:loopex}
\end{equation}
for a total initial four momentum $P$.
 The regularization of the loop function $G$, which is logarithmically divergent, has been usually done in the chiral unitary approach by means of dimensional regularization or with a three-momentum cutoff, $q_{\textrm{max}}$, and both usually provide equivalent results. However, 
it was justified in Refs.~\cite{wu,Xiao:2013jla,Ozpineci:2013qza}  
that the cutoff method is more convenient in the heavy flavor sector and, therefore, this is the regularization method we will use in the present work.
In terms of a three-momentum cutoff, the loop function reads
\begin{equation}
\label{eq:Gcutoff}
G=\int\limits_{0}^{q_{\textrm{max}}}\frac{d^3q}{(2\pi)^3}\,\frac{\omega_D+\omega_B}{2\omega_D\omega_B}\,\frac{1}{(P^0)^2-(\omega_D+\omega_B)^2+i\epsilon}\ ,
\end{equation}
with $\omega_{D(B)}=\sqrt{m_{D(B)}^2+\vec{q}^{\ 2}}$.
It was shown in ref.~\cite{Lu:2014ina} that in order to respect heavy quark symmetry in the unitarized hadron-hadron interaction a special $G$ function could be used which, however, was equivalent to stating that in the cutoff method the same cutoff, independent of heavy flavor, should be used. The same conclusion, with different arguments, was reached in  \cite{Ozpineci:2013qza}.
Hence we use values of the same order as those used in $B\bar B$ \cite{Ozpineci:2013qza}. Therefore, we will consider values $q_{\textrm{max}}\in[400,600]\mev$, where the differences in the results by varying the cutoff within this range can be considered as an estimation of the uncertainty in our calculation.
In Eq.~(\ref{eq:tI0swave}), $V$ is factorized out of the loop function since the momentum in the propagator of the exchanged vector  meson is neglected. However the running momentum inside the loop can reach values comparable to the exchanged vector meson mass. In Ref.~\cite{Xiao:2013jla} it was justified that this effect can be taken into account by including a factor $f^2(\vec q)$ in the integrand of Eq.~(\ref{eq:Gcutoff}), where $f(\vec q)$ is the form factor
\be
f(\vec q)=\frac{m_V^2}{\vec q\,^2+m_V^2}.
\label{eq:formfactsimple}
\ee
The factor corresponds to the propagator of the exchanged vector
neglecting the energy exchange $q^0$, which is zero in the on shell
diagonal transitions, $BD\rightarrow BD$ for instance, and we also take
it zero in the propagator of the exchanged vector in the loops, following the on shell factorization of the potential
as discussed in Refs.~\cite{Oller:1998zr,Oller:2000fj}.
Eq.~(\ref{eq:formfactsimple}) has assumed the external meson momentum to be zero. We can improve upon that, by considering an average initial momentum of the order of $p=\sqrt{2\mu B}$ where $B$ is the binding energy of the molecule and $\mu$ the reduced mass of its two components.
 Then $\vec q$ in Eq.~(\ref{eq:formfactsimple}) has to be replaced by $(\vec p-\vec q)$. After projecting over s-wave, for the new Eq.~(\ref{eq:formfactsimple}) we obtain the factor
 \be
 \tilde f(q)=\frac{m_V^2}{4 p q} 
 \ln\left[\frac{(p+q)^2+m_V^2}{(p-q)^2+m_V^2}\right].
 \ee
Note that this factor is never singular because $p$ is real, as it corresponds to an average over the momentum distribution of the molecular state.

On the other hand, in the evaluation of the $B^*D(\bar D)$, $BD^*(\bar D^*)$ and $B^*D^*(\bar D^*)$ interaction, there are vector mesons in the loop function whose polarization vectors should be carefully treated in the resummation implicit in the unitarization procedure. For the general vector-pseudoscalar interaction this was done in Ref.~\cite{Roca:2005nm}, where it was shown that, using the $\vec \epsilon\cdot\vec \epsilon\,'$ structure in the potential, the same Bethe-Salpeter equation (\ref{eq:T})  factorizing $\vec \epsilon\cdot\vec \epsilon\,'$ can be used, up to a correction in the loop function of $\vec q\,^2/(3m_V^2)$ which we can safely neglect.
Furthermore, the masses in the loop function must be changed to 
$m_{D^*}$ and/or $m_{B^*}$ accordingly for the corresponding channels.

\subsection{Results}
Since we are evaluating the interaction in s-wave, the possible quantum numbers of the different channels are, $J^P=0^+$ for $BD$; $1^+$ for
$B^*D$ and $BD^*$ and degenerate $0^+$, $1^+$, $2^+$ for $B^*D^*$. 
(All in isospin $I=0$ as explained below  Eq.~(\ref{eq:vfields})).
For  $B\bar D$,
$B^*\bar D$, $B\bar D^*$ and $B^*\bar D^*$ the spin-parities are the same as for the $B(B^*)D(D^*)$ case but now the isospin can be 0 or 1 (see Eqs.~(\ref{eq:tI0swavebar})
and (\ref{eq:tI1swavebar})).

\begin{table}[ht]
\begin{center}
\begin{tabular}{|c|c|c|c|c|c|}
\hline 
 & $I(J^P)$ & $\sqrt{s_p}$ & $B$ & $g$  &   $a$  [fm]\\
\hline
$BD$     & $0(0^+)$        &  7133$|$7111 & 15$|$38  &  33484$|$ 49867 	 &	-1.78$|$-1.45 \\ \hline
$B^*D$   & $0(1^+)$	   &  7179$|$7156 & 15$|$38  &  33742$|$ 50243  &	-1.78$|$-1.45 \\ \hline
$BD^*$   & $0(1^+)$  	   &  7270$|$7247 & 16$|$39  &  35171$|$ 52262	&	-1.75$|$-1.45 \\ \hline
$B^*D^*$ & $0(0^+,1^+,2^+)$&  7316$|$7293 & 16$|$39  &  35438$|$ 52652 	&	-1.75$|$-1.45 \\ \hline
$B\bar D$     & $0(0^+)$   &  7146$|$7140 & 1.7$|$8.4  &  13225$|$ 23296&	-3.77$|$-1.93 \\ \hline
$B^*\bar D$   & $0(1^+)$   &  7192$|$7186 & 1.7$|$8.4 &  13357$|$ 23494 & 	-3.74$|$-1.93 \\ \hline
$B\bar D^*$   & $0(1^+)$   &  7284$|$7277 & 2.1$|$9.5  &  14539$|$ 24915 & 	-3.32$|$-1.83 \\ \hline
$B^*\bar D^*$ & $0(0^+,1^+,2^+)$& 7330$|$7322 & 2.1$|$9.5 &  14678$|$ 25123 &	-3.31$|$-1.83 \\ \hline
$B\bar D$     & $1(0^+)$        &  -- & --  &  -- 			&	-0.53$|$-0.46 \\ \hline
$B^*\bar D$   & $1(1^+)$	   &  -- & --  &  -- 			&	-0.53$|$-0.46 \\ \hline
$B\bar D^*$   & $1(1^+)$  	   &  -- & --  &  -- 			&	-0.55$|$-0.46 \\ \hline
$B^*\bar D^*$ & $1(0^+,1^+,2^+)$&  -- & --  &  --			&	-0.55$|$-0.47 \\ \hline
\end{tabular}
\caption{Positions of the bound states ($\sqrt{s_p}$), binding energies ($B$) and couplings ($g$) of the different channels.
The first number in the last four columns represents the result for $q_{\textrm{max}}=400\mev$ and the second for $600\mev$.
 All units are in MeV except the scattering lengths, $a$, which are in fm.}
\label{tab:results}
\end{center}
\end{table}

By looking for poles in the second Riemann sheet of the unitarized
amplitudes, Eq.~(\ref{eq:T}), for $Re\{\sqrt{s}\}$ above the threshold
or in the physical sheet below, we  can see whether the interaction is
strong enough to generate dynamically a resonance in the former case or
a bound state in the latter one. In the present case,  for all the
channels {with $I=0$,} we find poles in the physical sheet below the
threshold which thus correspond to bound states. The position of
these
poles, $\sqrt{s_p}$, can be identified as the mass of the generated
bound state and are shown  in table \ref{tab:results} for different
values of the regularization cutoff, the first number in the last four
columns stands for the result for $q_{\textrm{max}}=400\mev$ and the
second one for $600\mev$. The difference in the values obtained for both
cutoffs should be regarded as the main uncertainty in our model. In the
table we also show  the corresponding binding energies
$B\equiv\sqrt{s_\textrm{threshold}}-\sqrt{s_p}$ and the values of the
couplings of the different poles to the corresponding channels, which
are defined considering that close to the pole
\be
T\simeq\frac{g^2}{s-s_p}\,,
\ee
 and they can be obtained by evaluating the residue of $T$ at the pole position.
  
 We find bound states, poles below threshold, for all the channels with $I=0$ at energies ranging a few hundred MeV above 7~GeV and binding energies of about 15-40~MeV for  $BD$, $B^*D$, $BD^*$ and $B^*D^*$ interactions and about 2-10~MeV for $B\bar D$, $B^*\bar D$, $B\bar D^*$ and $B^* \bar D^*$. For the latter channels, in $I=1$ the interaction is repulsive and, thus, no poles are found. Note that, despite the large uncertainty in the binding energy, stemming from the cutoff dependence, the absolute size of the binding energy is small compared to the mass of the system and is of the same order of magnitude as in other heavy flavor systems \cite{Xiao:2013yca,Dias:2014pva,Ozpineci:2013qza,Xiao:2013jla}. 
Note also that the binding energies are almost degenerate for all the channels. This is a manifestation of the independence of the binding energy on the heavy quark mass as a consequence of the HQSS \cite{Altenbuchinger:2013vwa,Lu:2014ina,Ozpineci:2013qza}.
For the $B\bar D$, $B^*\bar D$, $B\bar D^*$ and $B^* \bar D^*$ channels in $I=0$ the binding energy is very small, therefore the claim of their correspondence to actual mesons should be taken cautiously since further refinements of the model could make the pole disappear. However, the fact that we find poles for all the range of the cutoff considered is  a point in favor of their actual existence.
 
 In the last column of table \ref{tab:results}  we also show the values of the s-wave scattering lengths
\be
a=-\frac{1}{8\pi \sqrt{s_\textrm{th}}} T(\sqrt{s_\textrm{th}}),
\ee 
 with $\sqrt{s_\textrm{th}}$ the energy of the corresponding threshold, (and where we have used the  scattering length sign convention $p \cot \delta=\frac{1}{a}+\frac{1}{2}r_0 p^2$).
 
 It is worth stressing that the dynamics used here for the interaction, based on the HGS approach, stems from vector exchange. One can see that the source of attraction from this source in systems of this type is much bigger than the one obtained from pion exchange, via two step processes like $BD\to B^* D^* \to BD$ \cite{lianghqss,Uchino:2015uha}. In view of this it is not surprising that in \cite{Manohar:1992nd} no bound state for the $BD$ system was found using one pion exchange. We would like to note here that the exchange of vector mesons has also been introduced in quark models with the name of extended chiral quark model,
\cite{Zhang:1994pp,Zhang:1997ny,Dai:2003dz,Huang:2004sj,Huang:2004ke}  
  and its effects have been found to be important.

  The states found in the present work could in practice correspond to
  actual resonances with a narrow width which would come from
  subdominant channels with thresholds below the pole positions. It is
  worth mentioning that, according to the particle data table
  (PDG) \cite{pdg}, no mesons with both charm and beauty (in addition to
  the $B_c^+(6275)(0^-)$ and the $B_c(2S)^+(6842)(0^-)$) have been
  experimentally discovered.
  It is also worth noting that the poles in $I=0$ for the $B\bar D$, $B^*\bar
  D$, $B\bar D^*$ and $B^* \bar D^*$ would correspond to exotic mesons
  since they would contain a $\bar b$ and $\bar c$ quark at the same
  time.
   The findings in the present work are an indication that there is
  still much room to improve  the so far scarce experimental evidence of
  mesons with charm and beauty which would help understand the dynamics
  of the heavy flavor sector.

\section{Summary and conclusions}

We have done a theoretical study of the $BD$, $B^*D$, $BD^*$, $B^*D^*$,
$B\bar D$, $B^*\bar D$, $B\bar D^*$ and $B^* \bar D^*$ interaction to
try to see the possible dynamical generation of mesons with both charm
and beauty flavors. We evaluate the interaction starting from a tree
level elementary process obtained from suitable extensions of the hidden
gauge symmetry Lagrangians to heavy flavor, compatible with the heavy quark
spin symmetry of QCD, in order to evaluate the dominant mechanisms with
a vector meson exchange. 
We made a derivation of the Lagrangians in the heavy sector based on the hypothesis of having the heavy quarks as spectators.
We find an attractive and sizable potential for
the interaction in isospin $I=0$ for all the interactions. These
potentials are used as the kernel of the unitarization procedure using
the techniques of the coupled channels unitary approach which only depends on one
free regularization parameter. The dependence on the model on this
parameter, a three-momentum cutoff, represents the main source of
uncertainty of the model. By looking for poles of the unitarized
amplitudes we find poles below the thresholds of the different channels
with $I=0$ which thus correspond to bound
states with  quantum numbers $J^P=0^+$ for $BD$; $1^+$ for
$B^*D$ and $BD^*$ and degenerate $0^+$, $1^+$, $2^+$ for $B^*D^*$, at
energies slightly above 7~GeV and with binding energies  of about
20-60~MeV.  Similarly, for the  $B\bar D$  ($0^+$), $B^*\bar D$ ($1^+$),
$B\bar D^*$ ($1^+$) and $B^*\bar D^*$
($0^+$, $1^+$, $2^+$) interaction we also find bound states in $I=0$
but the interaction is repulsive in $I=1$. These latter bound states
would correspond to exotic mesons with $\bar b$ and $\bar c$ quarks.

{We find several states of the type $B_c$ which do not correspond to
the only two $B_c$ states so far reported in the PDG as the ground state
$B_c$ and $B_c(2S)$.
They are predictions that find an analogy with many states already
found in the $D_s$ sector.
On the other hand, we also find six new states of
$B(B^*)\bar{D}(\bar{D}^*)$ type, with $I=0$,
which are clearly exotic since they contain a $\bar{b}\bar{c}$ pair of
heavy quark, and are not of the $q\bar{q}$ type.
The results obtained here and the similarity of the states found to some
already observed in the $D_s$ states should stimulate the experimental
search of these states that should shed valuable light on hadron
dynamics.}

\section*{Acknowledgments}
This work is partly supported by the Spanish Ministerio
de Economia y Competitividad and European FEDER funds
under the contract number FIS2011-28853-C02-01, FIS2011-
28853-C02-02, FIS2014-57026-REDT, FIS2014-51948-C2-
1-P, and FIS2014-51948-C2-2-P, and the Generalitat Valenciana
in the program Prometeo II-2014/068 (EO).

\end{document}